\documentclass[11pt]{article}
\usepackage[margin=1in]{geometry}
\usepackage{anyfontsize,colortbl,xcolor}
\usepackage{bm,type1cm,verbatim,graphicx,latexsym,lscape}
\usepackage{caption,subcaption,booktabs}
\usepackage{setspace,color,bbm,amsfonts,amsmath}
\usepackage{epsfig,steinmetz}
\usepackage{hyperref}
\usepackage[numbers]{natbib}
\usepackage{amsthm}
\renewcommand\thetable{\arabic{table}}

\graphicspath{{./images/}}

\newtheorem{theorem}{Theorem}

\renewcommand\thetable{\arabic{table}}

\usepackage{epsfig}
\usepackage{amsmath}

\def\bSig\mathbf{\Sigma}

\newcommand{\bbo}{\mathbbm 1}

\title{A semiparametric approach for the estimation of covariate-adjusted area under the receiver operating characteristic curve}
\author{Shanshan Liu$^{1,\dagger}$ \and Jianlei Huang$^{2,\dagger}$ \and Guoqing Diao$^{1,*}$}
\date{}

\begin{document}
\maketitle
\begin{center}
\small
$^1$Department of Biostatistics and Bioinformatics, The George Washington University, Washington, DC, U.S.A.\\
$^2$Department of Mathematics and Statistics, Georgetown University, Washington, DC, U.S.A.\\[4pt]
$^\dagger$These authors contributed equally to this work.\\
$^*$Correspondence: Guoqing Diao (\texttt{gdiao@gwu.edu})
\end{center}

\begin{abstract}
Receiver operating characteristic (ROC) and the area under the ROC curve (AUC) are widely used to evaluate the discriminative ability of biomarkers. In many clinical settings, however, diagnostic accuracy varies substantially across patient characteristics, and failure to account for such heterogeneity can lead to misleading conclusions. We propose a new semiparametric framework based on generalized additive models to estimate covariate-specific and covariate-adjusted AUC while allowing for nonlinear and interaction effects of covariates on biomarker performance. Our method accommodates both binary and multicategory disease status. We also establish the asymptotic properties of the proposed estimators. Simulations demonstrate favorable finite-sample performance. We illustrate the method using data from the Alzheimer's Disease Neuroimaging Initiative, where substantial heterogeneity in biomarker discrimination across covariates is observed.
\end{abstract}

\medskip
\noindent\textbf{Keywords:} covariate adjustment; generalized additive model; restricted maximum likelihood; ROC analysis

\maketitle


\section{Introduction}
\label{sec:intro}
In medical diagnostics, the receiver operating characteristic (ROC) curve is a useful tool in statistical procedures such as classification and discrimination. For example, it is often used to evaluate the ability of a diagnostic test or biomarker to distinguish between different disease statuses. In particular, in the case of two disease statuses (e.g., healthy and diseased), the area under the ROC curve (AUC), which measures the probability that a diseased object gets a higher test result than a healthy one, is a classical tool to measure the accuracy of the test. When there are three or more statuses (e.g., healthy, intermediate, and diseased), these concepts can be generalized to the ROC surface and the volume under the ROC surface (VUS) naturally. In ROC analysis, if we believe that some covariate affects our test result, we may use this covariate to improve the accuracy of our measurement. This procedure is called covariate adjustment. Covariate-specific ROC ($ROC(\bm x)$) and AUC ($AUC(\bm x)$), which evaluate ROC and AUC when the covariate takes a specific value, are commonly used in the literature to examine the accuracy of a diagnostic test in the presence of covariates. In Janes and Pepe\cite{janes2009adjusting}, a new method for covariate-adjusted ROC curve (AROC) and AUC (AAUC) is proposed, which integrates $ROC(\bm x)$ and $AUC(\bm x)$, respectively, over the distribution of the covariate $\bm X$. AAUC summarizes the overall accuracy of the test in one single number while accounting for covariate effects. 

As reviewed by Pardo-Fernández et al.\cite{pardo2014review}, there are three approaches to deal with possible covariate effects on the ROC curve and AUC, which are based on conditional distribution, induced-regression, and direct-regression, respectively. As for the first approach \citep{lopez2008nonparametric,inacio2013bayesian,inacio2022covariate,martinez2025semiparametric}, two conditional distributions of the test result by diseased status (i.e., healthy or diseased) given the covariate values are assumed and used to recover $AUC(\bm x)$. The second approach \citep{pepe2003statistical,faraggi2003adjusting,yao2010nonparametric,gonzalez2011roc,hammouri2025rocanalysiscovariateadjustment} specifies two independent location-scale models, and $ROC(\bm x)$ and $AUC(\bm x)$ are derived by using the induced form. The third approach \citep{alonzo2002distribution,cai2002semiparametric,pepe2003statistical,dodd2003semiparametric,rodriguez2011new,fanjulhevia2024newtestassessingcovariate} uses a regression model to describe the effect of covariates on ROC and AUC directly.

In the literature, much attention has been paid to correcting the bias in estimated AUC in the presence of covariate effects. However, to our knowledge, how the interactions between covariates would affect the result of ROC analysis has not been sufficiently discussed. Although some existing methods \citep{gonzalez2011roc,to2022estimation} consider a nonparametric location-scale model, they treat the covariates as a $p$-dimensional vector without explicitly modeling interactions among them. In kernel methods \citep{gonzalez2011roc} and local regression models \citep{to2022estimation}, such interactions are often too weak to capture or treated as noise, especially when the sample size is small or the assumptions in the GEE-based estimators are violat. Generalized additive models (GAMs), on the other hand,  provide a flexible compromise between parametric and fully nonparametric approaches. They extend generalized linear models by allowing model parameters—such as the conditional mean and, in more general formulations, additional distributional characteristics—to depend on smooth functions of the covariates while retaining a structured additive form \citep{hastie1986generalized,wood2017generalized}. The use of penalized smoothers stabilizes estimation and avoids the high variability often associated with fully nonparametric methods in moderate sample sizes, while preserving substantial flexibility to capture complex relationships. This flexibility is particularly valuable for AUC estimation, as discrimination measures can be sensitive to distributional features beyond simple location–scale differences. Moreover, GAMs readily accommodate nonlinear effects and interactions among covariates in an interpretable manner. Motivated by these advantages, we propose GAM-based estimators for covariate-specific AUC and adjusted AUC (AAUC), obtained via penalized likelihood with smoothing parameters selected by restricted maximum likelihood (REML) \citep{wood2016smoothing,wood2017generalized} and computed using an efficient Newton-type algorithm.

The rest of the paper is organized as follows. In Section 2, we propose GAM-based semi-parametric estimators of $AUC(\bm x)$, $VUS(\bm x)$, as well as AAUC and covariate-adjusted VUS (AVUS) in the presence of interactions. In Section 3, theoretical results regarding the asymptotic behavior of our estimators are presented. Several simulation studies are presented in Section 4, followed by a real example from the Alzheimer’s disease research in Section 5. Finally, we make the concluding remarks in Section 6.

\section{Methods}
\subsection{Notations and Target Parameters}

For each subject we observe the triplet $(Y,D,\bm X)$, where $Y$ denotes the result of the diagnostic test or biomarker under evaluation, $D$ denotes the true disease status, and $\bm X=(X_1,\ldots,X_p)^{T}$ is a $p\times 1$ vector of covariates with support $\mathcal X\subseteq\mathbb R^{p}$. We first develop the framework for a binary disease status $D\in\{0,1\}$, where $D=0$ and $D=1$ index the non-diseased and diseased populations, respectively; the extension to $M\ge 3$ ordered classes is given at the end of this subsection. The observed data consist of $n$ independent replicates $(Y_i,D_i,\bm X_i)$, $i=1,\ldots,n$, and we write $n_k=\#\{i:D_i=k\}$ for the size of disease group $k$, so that $\sum_k n_k=n$.

Throughout this subsection we impose no structure on the manner in which the test result depends on the covariates. We assume only that, within each disease group, $Y$ given $\bm X$ exists and is continuous. Accordingly, for $k=0,1$ we let
\begin{equation}\label{eq:cond-cdf}
F_k(y\mid\bm x)=P(Y\le y\mid D=k,\ \bm X=\bm x)
\end{equation}
denote the conditional cumulative distribution function (c.d.f.) of the test result in group $k$ at covariate value $\bm x$, and we write $f_k(y\mid\bm x)$ for the corresponding conditional density. The disease groups are modeled separately, so that $F_0$ and $F_1$ are unrestricted relative to one another.

Let $H$ denote the joint c.d.f. of the covariate vector $\bm X$ in the population of interest. For a fixed $\bm x\in\mathcal X$, let $Y^{(0)}$ and $Y^{(1)}$ be independent random variables with distributions $F_0(\cdot\mid\bm x)$ and $F_1(\cdot\mid\bm x)$. Without loss of generality, we assume that the stochastic ordering of the test results is $Y^{(1)}>Y^{(0)}$, given $X=x$, which means that higher test result indicates higher probability for disease at a fix covariate point. In Li et al.\cite{li2014sorting}, unknown stochastic ordering is discussed. 

The covariate-specific AUC at covariate value $\bm x$ is defined as the probability that a diseased subject yields a higher test result than a non-diseased subject with the same covariate value, 
\begin{equation}\label{eq:auc-x}
\mathrm{AUC}(\bm x)
=P\{Y^{(1)}>Y^{(0)}\mid \bm X=\bm x\}.
\end{equation}
Averaging \eqref{eq:auc-x} over the covariate distribution yields the covariate-adjusted AUC \citep{liu2013roc}, 
\begin{equation}\label{eq:aauc}
\mathrm{AAUC}=\int_{\mathcal X}\mathrm{AUC}(\bm x)\,dH(\bm x),
\end{equation}
which summarises the overall discriminatory ability of the test.

Expression \eqref{eq:auc-x} is the covariate-specific analogue of the Mann--Whitney (MW) representation of the AUC. When no covariates are present, $F_k(\cdot\mid\bm x)$ reduces to the marginal c.d.f. $F_k(\cdot)$, and substituting the empirical c.d.f.s of the two groups into \eqref{eq:auc-x} returns the familiar two-sample MW statistic $(n_0n_1)^{-1}\sum_{i:D_i=0}\sum_{j:D_j=1}\bbo(Y_j>Y_i)$, the nonparametric estimator studied by Nakas et al.\citet{nakas2004ordered} In the presence of covariates, since in general no two subjects share the same covariate value and the group-specific empirical distributions at $\bm X=\bm x$ are unavailable. Estimation of the conditional distributions $F_0(\cdot\mid\bm x)$ and $F_1(\cdot\mid\bm x)$ is therefore required. This is discussed in Section 2.2.

We next consider the case of three or more disease status. For the three-class case we index the classes by $D\in\{1,2,3\}$ in order of increasing disease severity, and for fixed $\bm x$ we let $Y^{(1)},Y^{(2)},Y^{(3)}$ be independent with distributions $F_1(\cdot\mid\bm x),F_2(\cdot\mid\bm x),F_3(\cdot\mid\bm x)$, respectively. The covariate-specific volume under the ROC surface (VUS) and the covariate-adjusted VUS are then,\begin{equation}\label{eq:vus-x}
\mathrm{VUS}(\bm x)
=P\{Y^{(1)}<Y^{(2)}<Y^{(3)}\mid\bm X=\bm x\}, \nonumber
\end{equation}
\begin{equation}\label{eq:avus}
\mathrm{AVUS}=\int_{\mathcal X}\mathrm{VUS}(\bm x)\,dH(\bm x)=E\{\mathrm{VUS}(\bm X)\}. \nonumber
\end{equation}

The definitions above depend on the data-generating mechanism only through the group-specific conditional distributions $F_k(\cdot\mid\bm x)$ and the covariate distribution $H$. They are therefore model-free, and any procedure that delivers estimates of $F_k(\cdot\mid\bm x)$ may be substituted into them to form the corresponding estimators.



\subsection{Distributional Generalized Additive Models for the Conditional Distributions}
We now introduce the working model used in this paper. We assume that, within disease group $k$, the conditional distribution of the test result belongs to a parametric family $\mathcal D$ indexed by $Q$ parameters, each of which is permitted to vary with the covariates:
\begin{equation}\label{eq:gam-model}
Y\mid D=k,\ \bm X=\bm x\ \sim\ \mathcal D\{\theta_{k1}(\bm x),\ldots,\theta_{kQ}(\bm x)\},
\qquad k=0,1,
\end{equation}
so that $F_k(y\mid\bm x)=F_{\mathcal D}\{y;\bm\theta_k(\bm x)\}$ with $\bm\theta_k(\bm x)=\{\theta_{k1}(\bm x),\ldots,\theta_{kQ}(\bm x)\}^{T}$, and $f_{\mathcal D}$ denotes the corresponding density. Each distributional parameter is linked to its own additive predictor,
\begin{equation}\label{eq:gam-predictor}
g_q\{\theta_{kq}(\bm x)\}=\eta_{kq}(\bm x)=\beta_{kq0}+\sum_{j=1}^{J_{kq}}f_{kqj}(\bm x),
\qquad q=1,\ldots,Q,
\end{equation}
where $g_q$ is a known link function relating the $q$th distributional parameter $\theta_q$ to the additive predictor $\eta_q$, , $\beta_{kq0}\in\mathbb R$ is an intercept, and each $f_{kqj}$ is an unknown smooth function depending on $\bm x$. Model \eqref{eq:gam-model}--\eqref{eq:gam-predictor} is a distributional generalized additive model, also known as a generalized additive model for location, scale and shape (GAM) \citep{stasinopoulos2024generalized,wood2017generalized}.

Interactions among covariates are accommodated by including smooth terms of more than one argument. In the two-covariate case $\bm x=(x_1,x_2)^{T}$ used in our derivations and simulations, we take
\begin{equation}\label{eq:gam-two-cov}
\eta_{kq}(\bm x)=\beta_{kq0}+f_{kq1}(x_1)+f_{kq2}(x_2)+f_{kq3}(x_1,x_2),
\end{equation}
where $f_{kq1}$ and $f_{kq2}$ are univariate smooth main effects and $f_{kq3}$ is a bivariate smooth interaction surface.

The exact form of \eqref{eq:gam-predictor} depend on the choice of distribution. For example, if the test results are assumed to be normally distributed, then $Q=2$ with $\theta_{k1}=\mu_k$ and $\theta_{k2}=\sigma_k$ denoting the conditional mean and standard deviation, $h_1$ is the identity link and $h_2$ is the logarithmic link, which ensures positivity of $\sigma_k(\bm x)$. Heavier-tailed alternatives, such as the $t$ family with an additional degrees-of-freedom parameter, are obtained by increasing $Q$ and are used in Sections 4 and 5.

For estimation, each smooth function is represented through a basis expansion, $f_{kqj}(\cdot)=\sum_{l=1}^{L_{kqj}}\gamma_{kqjl}\,b_{kqjl}(\cdot)$, where the $b_{kqjl}$ are known basis functions and $\bm\gamma_{kqj}=(\gamma_{kqj1},\ldots,\gamma_{kqjL_{kqj}})^{T}$ are unknown coefficients. Smoothness is enforced by the working prior $\bm\gamma_{kqj}\sim\mathcal N(\bm 0,\tau_{kqj}^{2}\bm K_{kqj}^{-})$, where $\bm K_{kqj}$ is a known penalty matrix determined by the basis together
with the roughness measure being penalized, and generally second-order penalties are used,  $\bm K_{kqj}^{-}$ denotes its generalized inverse, and $\tau_{kqj}^{2}$ governs the degree of smoothing, as implemented in In the \texttt{gamlss} package \citep{stasinopoulos2024generalized}.

Because the disease groups are modelled separately, estimation is carried out independently within each group $k$ by maximizing the penalized likelihood, which is solved by Newton method through iterative reweighted least squares (IRLS)
\begin{equation}\label{eq:penlik}
\ell_{p}^{(k)}
=\sum_{i:D_i=k}\log f_{\mathcal D}\{Y_i;\bm\theta_k(\bm X_i)\}
-\frac12\sum_{q=1}^{Q}\sum_{j=1}^{J_{kq}}\lambda_{kqj}\,
\bm\gamma_{kqj}^{T}\bm K_{kqj}\bm\gamma_{kqj},
\end{equation}
where $\lambda_{kqj}$ are smoothing parameters associated with each penalty. The quadratic penalty discourages excessive wiggliness of the smooth functions and is equivalent to treating the spline coefficients as random effects in a mixed-model representation.

The restricted maximum likelihood criterion (REML) is commonly used to estimate the smoothing parameters $\lambda_{kqj}$ (or equivalently the variance components $\tau_{kqj}^{2}$). REML accounts for the uncertainty in estimating fixed effects and typically yields more stable smoothing than criteria based on prediction error such as generalized cross-validation. In practice, estimation alternates between updating the model coefficients via penalized IRLS and updating the smoothing parameters via REML until convergence.

Details on Newton-type algorithms for fitting smooth functions and the REML criterion are described elsewhere\citet{wood2016smoothing,wood2004stable,wood2017generalized}. This method described above is implemented in standard software such as the \texttt{gamlss} \citep{gamlss} and \texttt{mgcv} \citep{R-mgcv} packages in $R$.


\subsection{Estimation and Inference}
Let $\widehat{\bm\theta}_k(\bm x)$ denote the estimator of $\bm\theta_k(\bm x)$ and let $\widehat{F}_k(\cdot|\bm{x})$ and $\widehat{f}_k(\cdot|\bm{x})$ be the CDF and density induced by the fitted GAM model, by plugging in the estimated parameters from the GAM estimation stage. We can obtain an estimator for the MW representation for covariate-specific AUC and VUS:
\begin{align}
\widehat{\mathrm{AUC}}(\bm x)&=\int_{-\infty}^{\infty}\widehat F_0(y\mid\bm x)\,\widehat f_1(y\mid\bm x)\,dy ,
\label{eq:auc-hat}\\
\widehat{\mathrm{VUS}}(\bm x)&=\int_{-\infty}^{\infty}\widehat F_1(y\mid\bm x)\{1-\widehat F_3(y\mid\bm x)\}\,\widehat f_2(y\mid\bm x)\,dy .
\label{eq:vus-hat}
\end{align}

It follows that we can obtain the covariate-adjusted measures by integrating \eqref{eq:auc-hat} and \eqref{eq:vus-hat} over the empirical distribution $\widehat H$ of the covariates, which places mass $n^{-1}$ at each observed $\bm X_i$:
\begin{equation}\label{eq:aauc-hat}
\widehat{\mathrm{AAUC}}=\int_{\mathcal X}\widehat{\mathrm{AUC}}(\bm x)\,d\widehat H(\bm x)=\frac1n\sum_{i=1}^{n}\widehat{\mathrm{AUC}}(\bm X_i),
\qquad
\widehat{\mathrm{AVUS}}=\frac1n\sum_{i=1}^{n}\widehat{\mathrm{VUS}}(\bm X_i).
\end{equation}

Note that here we assume the underlying distribution family for $Y$ given $D$ and $\bm X$ is known, e.g., Gaussian or t-distribution, while in practice it may be selected through cross validation. In the Gaussian case, we have a closed-form after simplification; for other distributions we can use numerical approximation method such as the adaptive gaussian quadrature, which is conveniently implemented by \texttt{integrate} in R.

For statistical inference, the sampling variability of the proposed estimators does not admit a simple closed-form expression due to the multi-stage estimation procedure involving distributional GAM fitting and numerical integration. We therefore estimate variances using the nonparametric bootstrap. Specifically, bootstrap samples are obtained by resampling the observations with replacement, refitting the GAM models, and recomputing the estimators for each sample. Standard errors are calculated as the empirical standard deviation of the bootstrap replicates, and confidence intervals are constructed using a normal approximation.

\section{Asymptotic Properties}

We establish the large-sample
behavior of the proposed estimators for covariate-specific and covariate-adjusted
ROC functionals stated as follow. 

\subsection{Covariate-specific accuracy measures}

\begin{theorem}
Under the assumption that the covariance matrix of $\bm X$ is finite and positive definite and additional assumptions (A1)-(A4) in the Appendix, for each fixed $\bm x \in \mathcal X$, we can show that 
\[
\sqrt{n}\{\widehat{\mathrm{AUC}}(\bm x)-\mathrm{AUC}(\bm x)\}
\Rightarrow N\{0,\sigma^2_{\mathrm{AUC}}(\bm x)\}.
\]
Similarly, 
\[
\sqrt{n}\{\widehat{\mathrm{VUS}}(\bm x)-\mathrm{VUS}(\bm x)\}
\Rightarrow N\{0,\sigma^2_{\mathrm{VUS}}(\bm x)\}.
\]
\end{theorem}

\subsection{Covariate-adjusted accuracy measures}

\begin{theorem}
Under the assumption that the covariance matrix of $\bm X$ is finite and positive definite and additional assumptions (A1)-(A4) in the Appendix, we can show that
\[
\sqrt{n}\{\widehat{\mathrm{AAUC}}-\mathrm{AAUC}\}
\Rightarrow N(0,\sigma^2_{\mathrm{AAUC}})
\]
and
\[
\sqrt{n}\{\widehat{\mathrm{AVUS}}-\mathrm{AVUS}\}
\Rightarrow N(0,\sigma^2_{\mathrm{AVUS}}).
\]
\end{theorem}

Here $\sigma^{2}_{\mathrm{AUC}}(\bm x)$ and
$\sigma^{2}_{\mathrm{VUS}}(\bm x)$ denote the asymptotic variances of the
covariate-specific estimators and $\sigma^{2}_{\mathrm{AAUC}}$ and
$\sigma^{2}_{\mathrm{AVUS}}$ those of the covariate-adjusted estimators; explicit expressions are given in the Appendix. The asymptotic variances depend on contributions from estimation of the conditional distributions within each disease group as well as from sampling variability in the covariate distribution. Inexplicit expressions are given in the Appendix. practice, standard errors are obtained via the nonparametric bootstrap as described in Section~2.3. Proofs are given in the Appendix and are based on the asymptotic properties of the distributional GAM estimators together with smoothness of the ROC functionals and integration over the empirical covariate distribution.

\section{Simulation Studies} 

\subsection{Simulation Setting}
We conduct simulation studies to examine the performance of the proposed GAM-based approaches for estimating covariate-specific AUC. In particular, we examine the Monte Carlo mean square errors (MSEs) of the estimators. In the simulation study, we set the following sample size: $n_0=n_1=50,100,200$, and we perform $1000$ Monte Carlo replications for each experiment.

We have the following setting for our experiments:
\begin{align*}
\mu_1(\bm{X})&=1+(X_1-0.5)+(1-0.5\times X_2+X_2^2)+(X_1\times \sin(\pi X_2)),\\
\text{and }\mu_0(\bm{X})&=0+0.8\times(X_1-0.5)+(e^{X_2})+(X_1\times(0.5+\sin(2.5 X_2)),
\end{align*}
where $X_1$ and $X_2$ are independently generated from a $U(0,1)$ distribution. Conditional on $D=k$, and $\bm X = \bm x$, the test result is generated as
\[
Y=\mu_k(\bm x) +\epsilon_k,k=1,0.
\]
The error terms $\epsilon_1$ and $\epsilon_0$ are independently generated in $4$ scenarios:
\begin{align*}
&1.\ \epsilon_1\sim N(0,1)\ \epsilon_0\sim N(0,1)\\
&2.\ \epsilon_1\sim N(0,1)\ \epsilon_0\sim N(0,2)\\
&3.\ \epsilon_1\sim t(3)\ \epsilon_0\sim t(5)\\
&4.\ \epsilon_1\sim N(0,1)\ \epsilon_0\sim t(5)
\end{align*} 
In the first scenario, $\epsilon_1$ and $\epsilon_0$ are draw from $i.i.d.$ distributions, while in latter three scenarios, they are not identically distributed.

\subsection{Comparison to existing Mann--Whitney Estimators}

In this section, we use the same simulation setting in To et al.\cite{to2022estimation} to examine the performance of the proposed estimator of VUS in presence of heteroscedascity and compare it to two existing estimators: (1) a Mann--Whitney estimator based on semiparametric generalized
estimating equations (MW-GEE), and (2) a Mann--Whitney estimator based on local linear regression (MW-LL). These scenarios involve three ordered disease classes, indexed by $D\in\{1,2,3\}$ as in Section~2.1, and a covariate that acts on both the location and the scale of the test result; they are therefore fitted with the distributional GAM of Section~2.2 taking $Q=2$, so that both $\eta_{k1}$ and $\eta_{k2}$ are smooth functions of the covariate. As in Section 4.1, we set the following sample size $n_1=n_2=n_3=50,100$ and $200$, and perform 500 Monte Carlo replications for each setting. We obtain 95\% confidence intervals based on 500 bootstrap samples. The simulation setting is as follows:
\[
Y=\mu_k(x)+\sigma_k^2(x)\epsilon_k,~~k=1,2,3,
\]
where $\sigma_k^2(\cdot)$ denotes the conditional variance function specified below.

Scenario 1: We generate $X \sim U(0,1)$ and $\epsilon_k \sim N(0,1)$ independently. The mean and variance functions are: 
\begin{align*}
\mu_1(x)&=1-0.5x+x^2\\
\mu_2(x)&=1.5+0.5x+2x^3\\
\mu_3(x)&=2+3x\\
\sigma^2_k(x)&=(1-x+x^2)^2 \ \ \ k=1,2,3.
\end{align*}
Scenario 2: We generate $X\sim U(0,1)$. The mean and variance functions are the same as in Scenario 1, while the error terms $\epsilon_1,\epsilon_2$ and $\epsilon_3$ are generated from $t$-distributions with degree of freedoms 5, 3 and 3, respectively, and then re-scaled to have unit variance: 
Scenario 3: We generate $X\sim U(0.5,1.5)$, and the error term $\epsilon_k\sim N(0,1)$ independently. The mean and variance functions are: 
\begin{align*}
\mu_1(x)&=-0.3+\sin(2\pi x)\\
\mu_2(x)&=1.5+\sin(2\pi x)\\
\mu_3(x)&=2+\sin(1.5x)\\
\sigma^2_j(x)&=(0.5+1.2x)^2 \ \ \ j=1,2,3.
\end{align*}

\begin{figure}[htbp]
\centering
\subfloat[Scenario I]{%
  \includegraphics[width=0.48\textwidth,keepaspectratio]{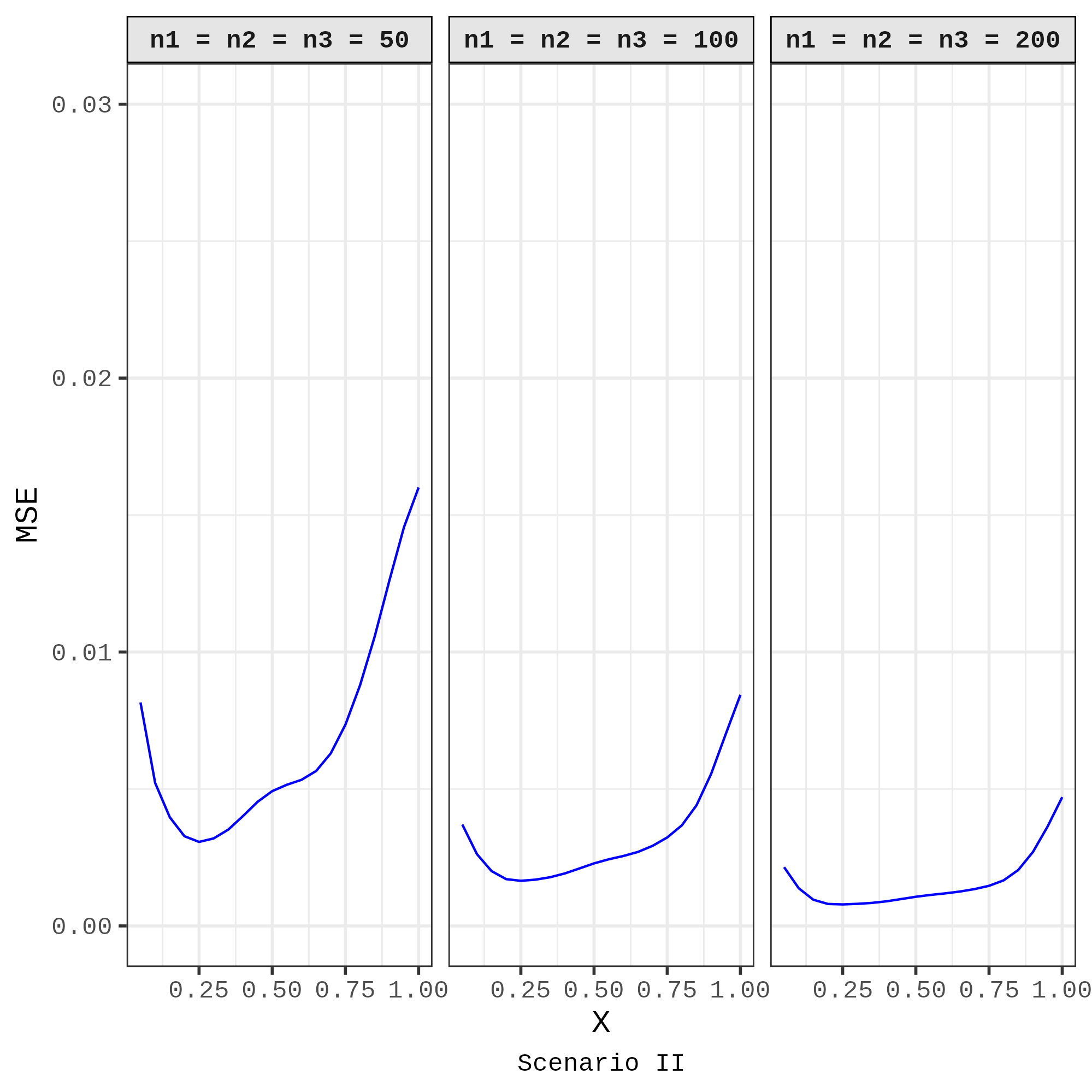}}
\hfill
\subfloat[Scenario II (bsnorm)]{%
  \includegraphics[width=0.48\textwidth,keepaspectratio]{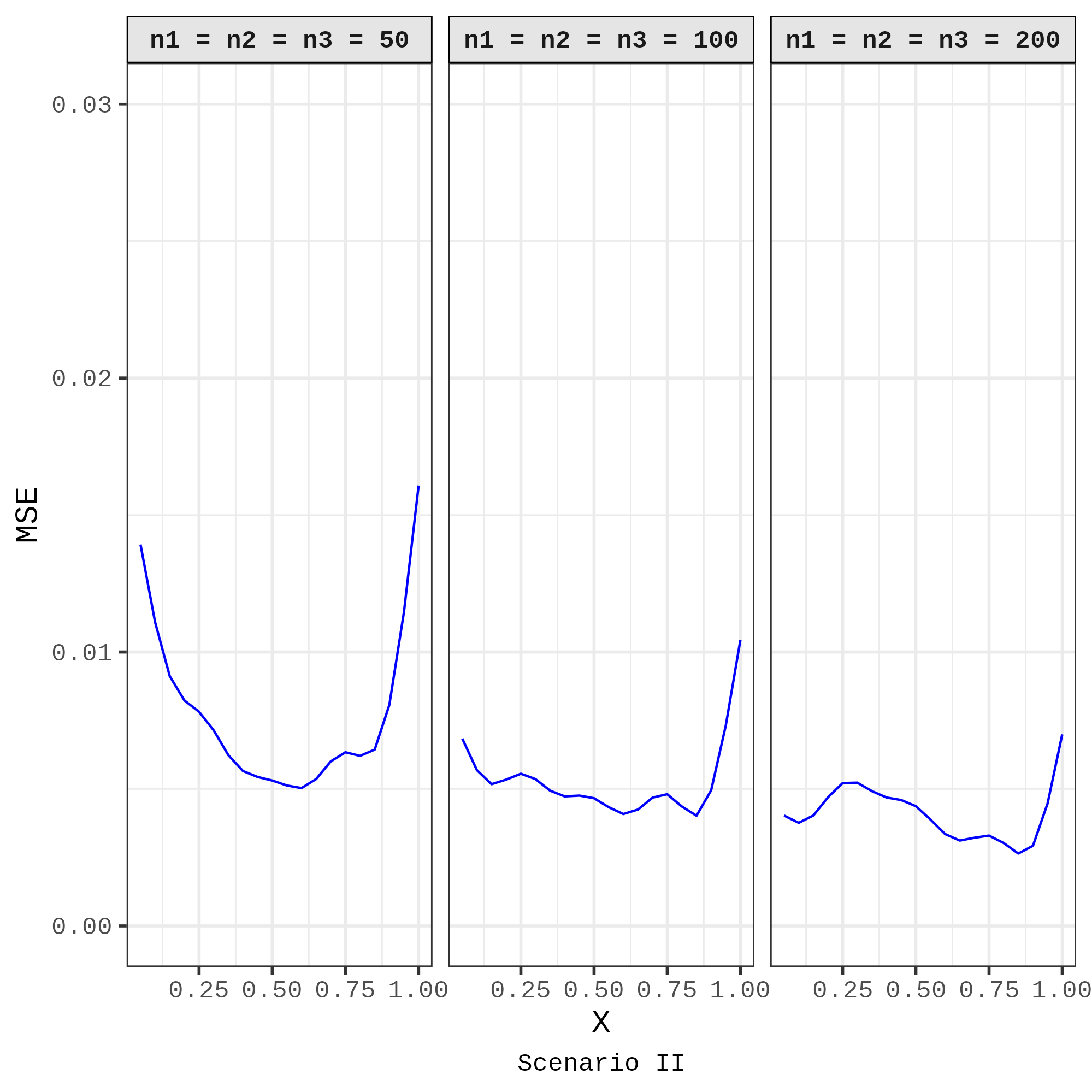}}
\\[0.8em]
\subfloat[Scenario II (gamlss)]{%
  \includegraphics[width=0.48\textwidth,keepaspectratio]{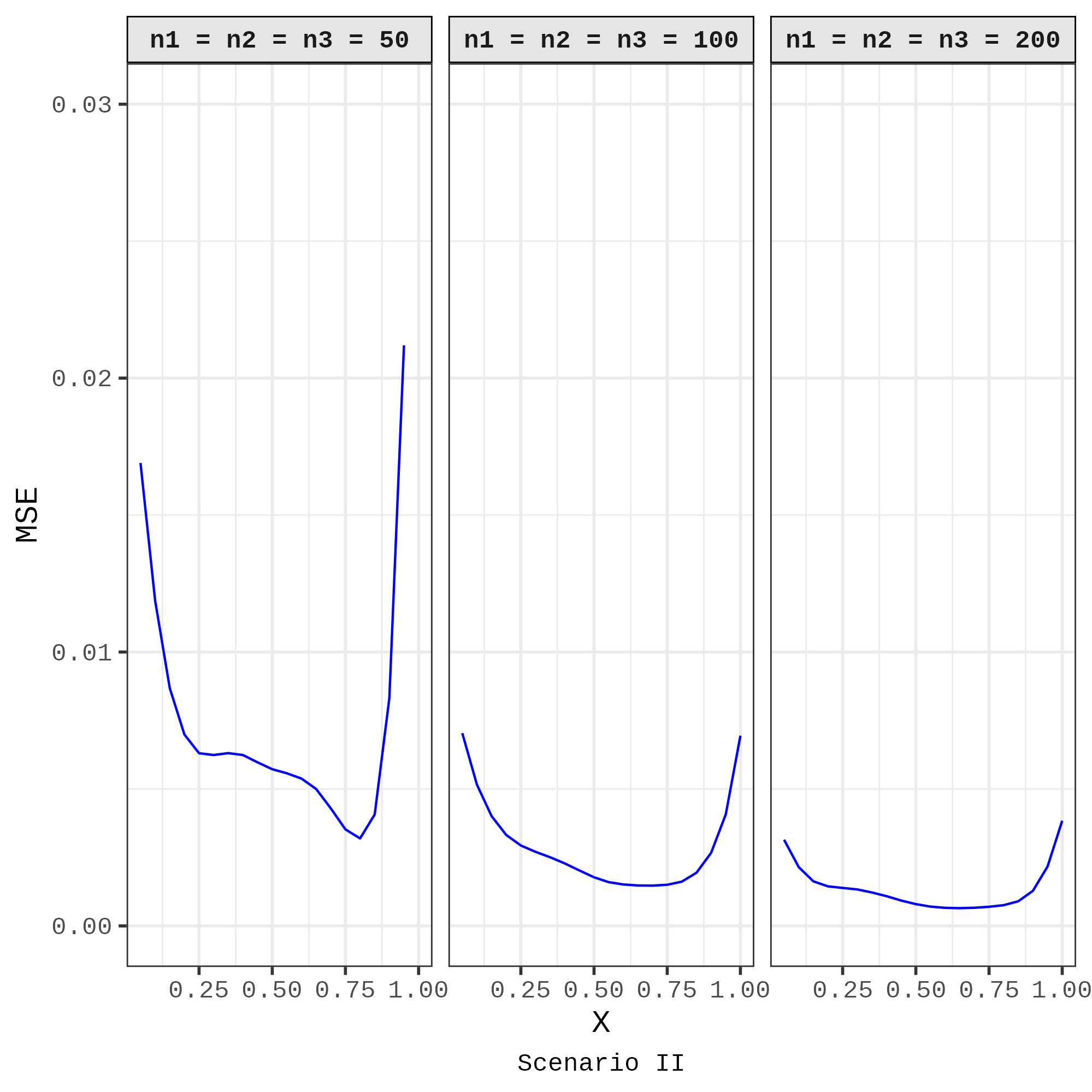}}
\hfill
\subfloat[Scenario III]{%
  \includegraphics[width=0.48\textwidth,keepaspectratio]{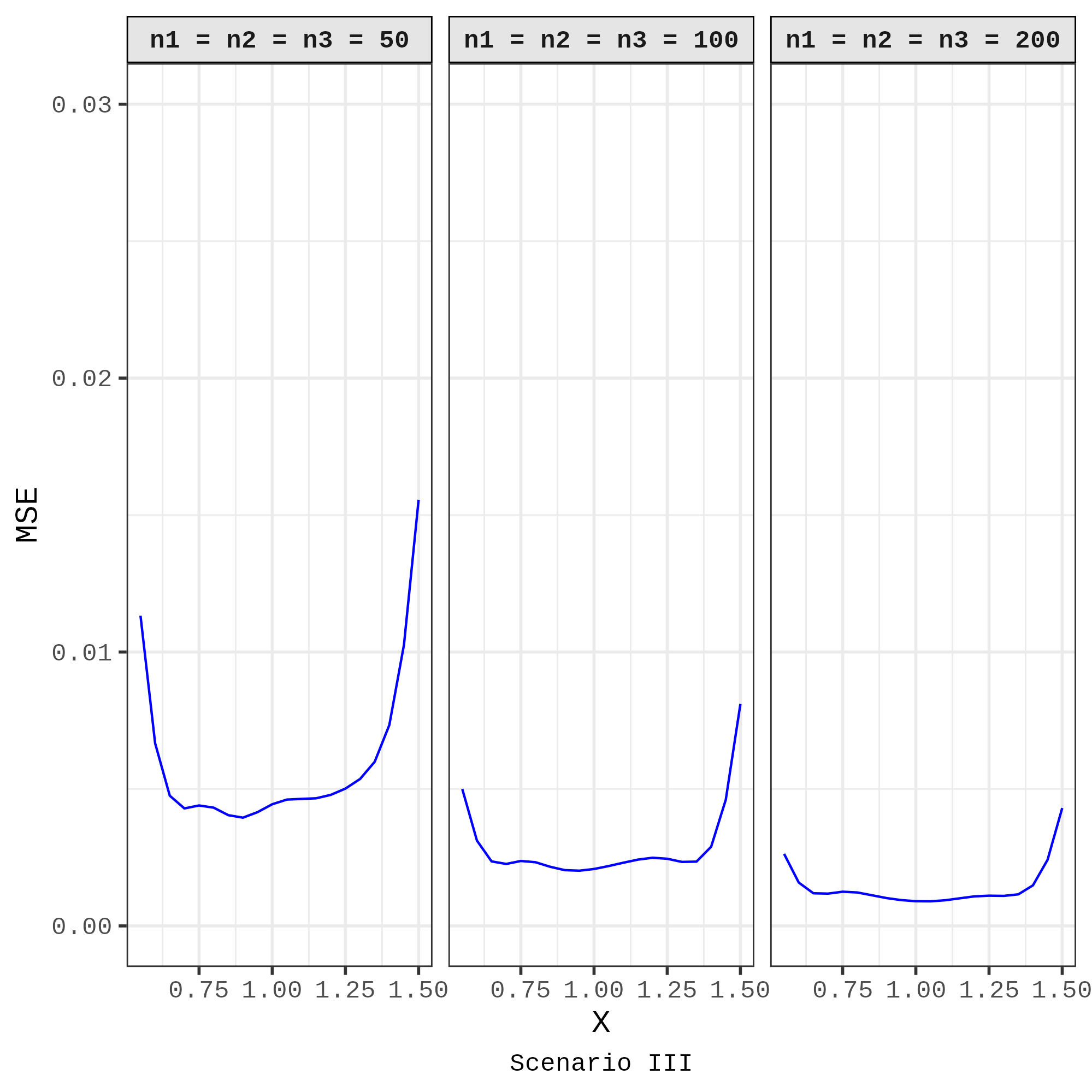}}
\caption{Mean square error (MSE) of GAM-based covariate-specific VUS in
scenario I, II and III.}\label{fig:vus-mse}
\end{figure}

\begin{figure}[htbp]
\centering
\subfloat[Scenario I]{%
  \includegraphics[width=0.48\textwidth,keepaspectratio]{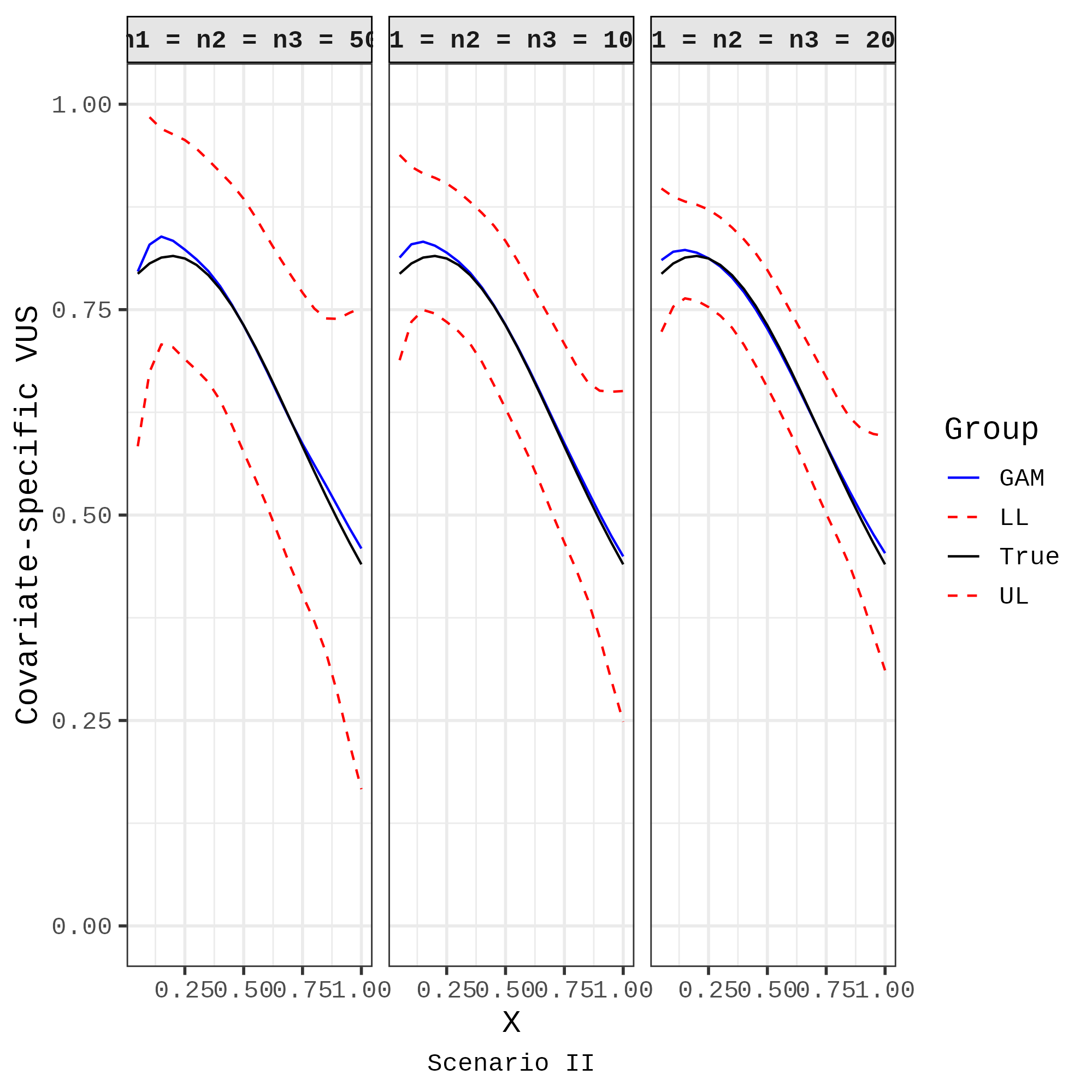}}
\hfill
\subfloat[Scenario II (bsnorm)]{%
  \includegraphics[width=0.48\textwidth,keepaspectratio]{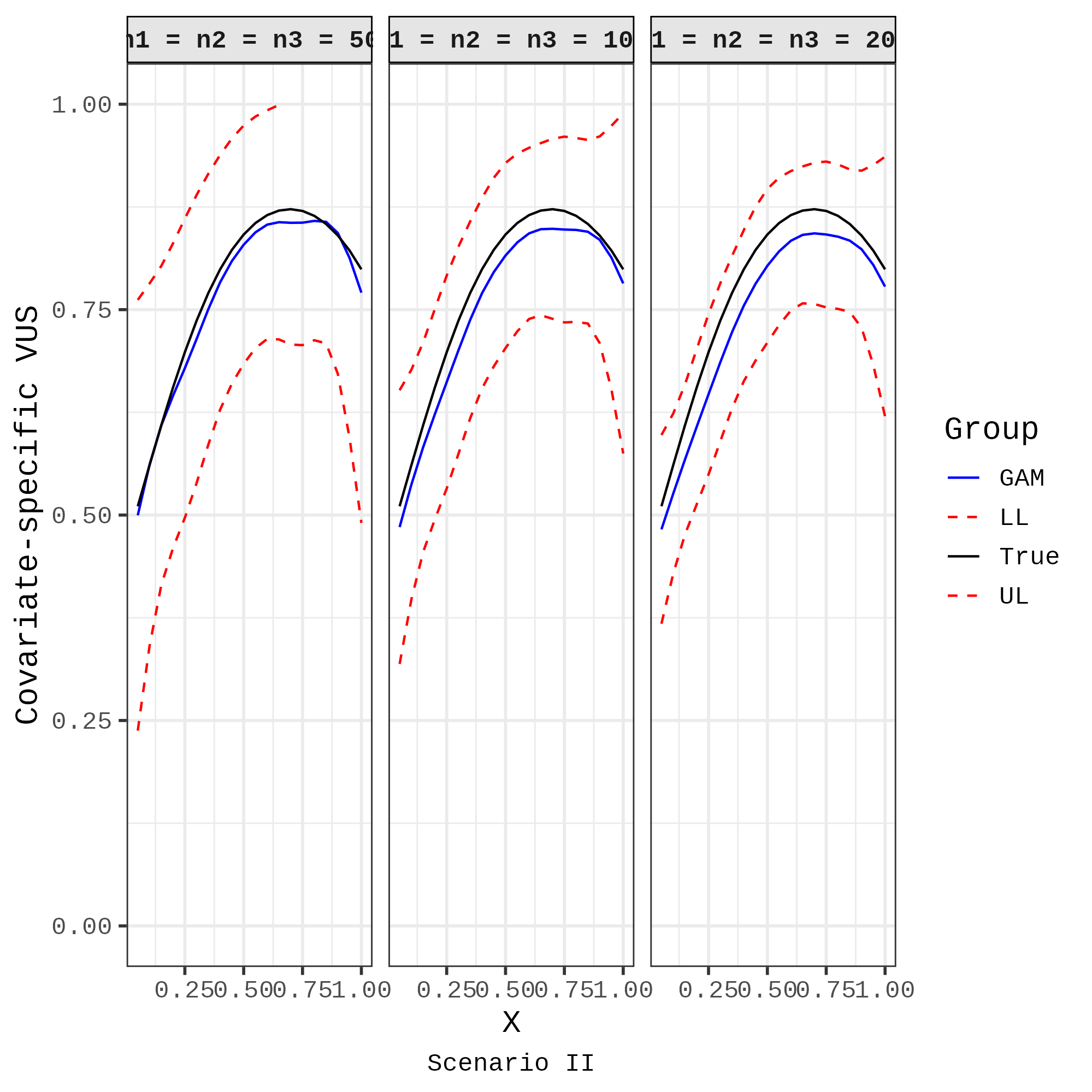}}
\\[0.8em]
\subfloat[Scenario II (gamlss)]{%
  \includegraphics[width=0.48\textwidth,keepaspectratio]{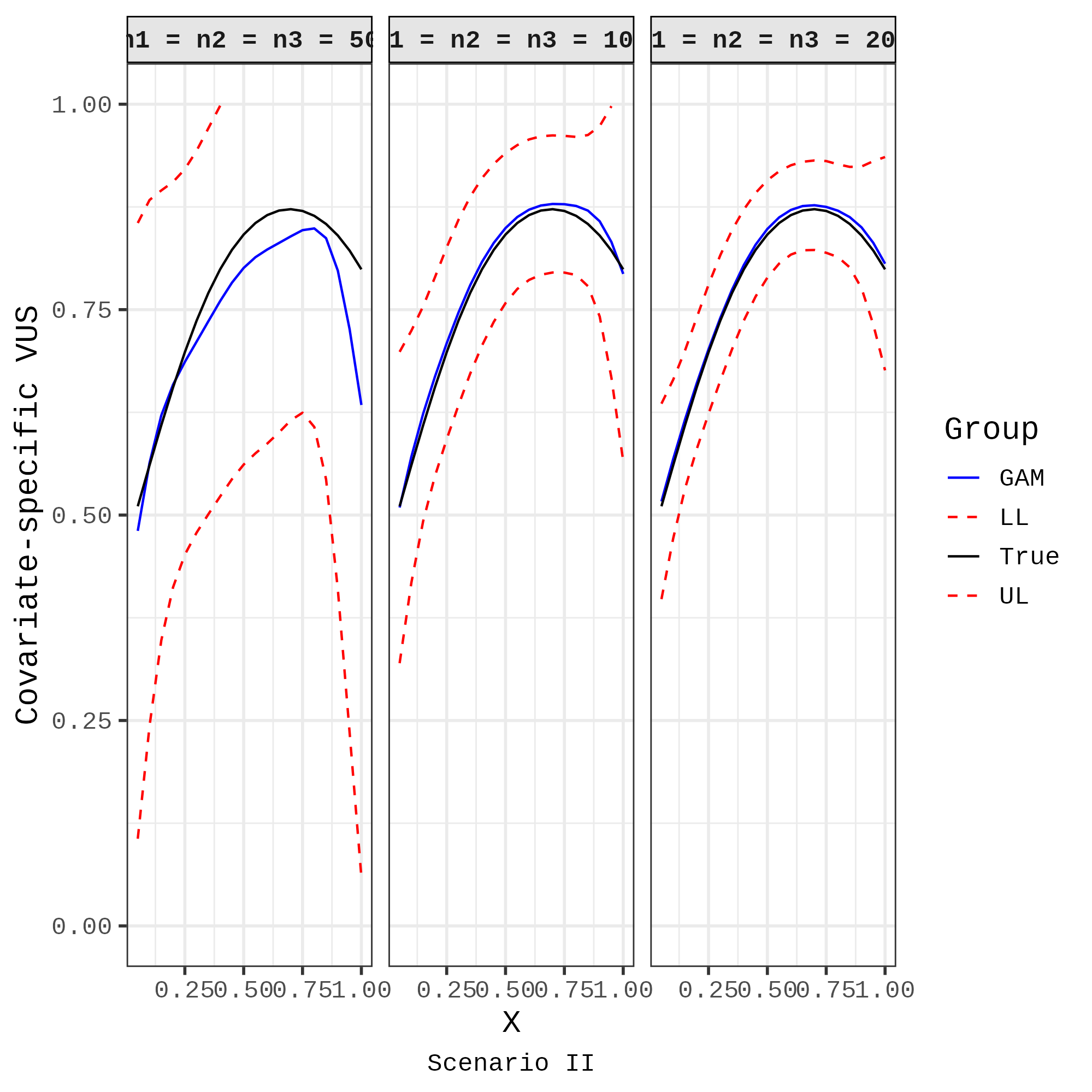}}
\hfill
\subfloat[Scenario III]{%
  \includegraphics[width=0.48\textwidth,keepaspectratio]{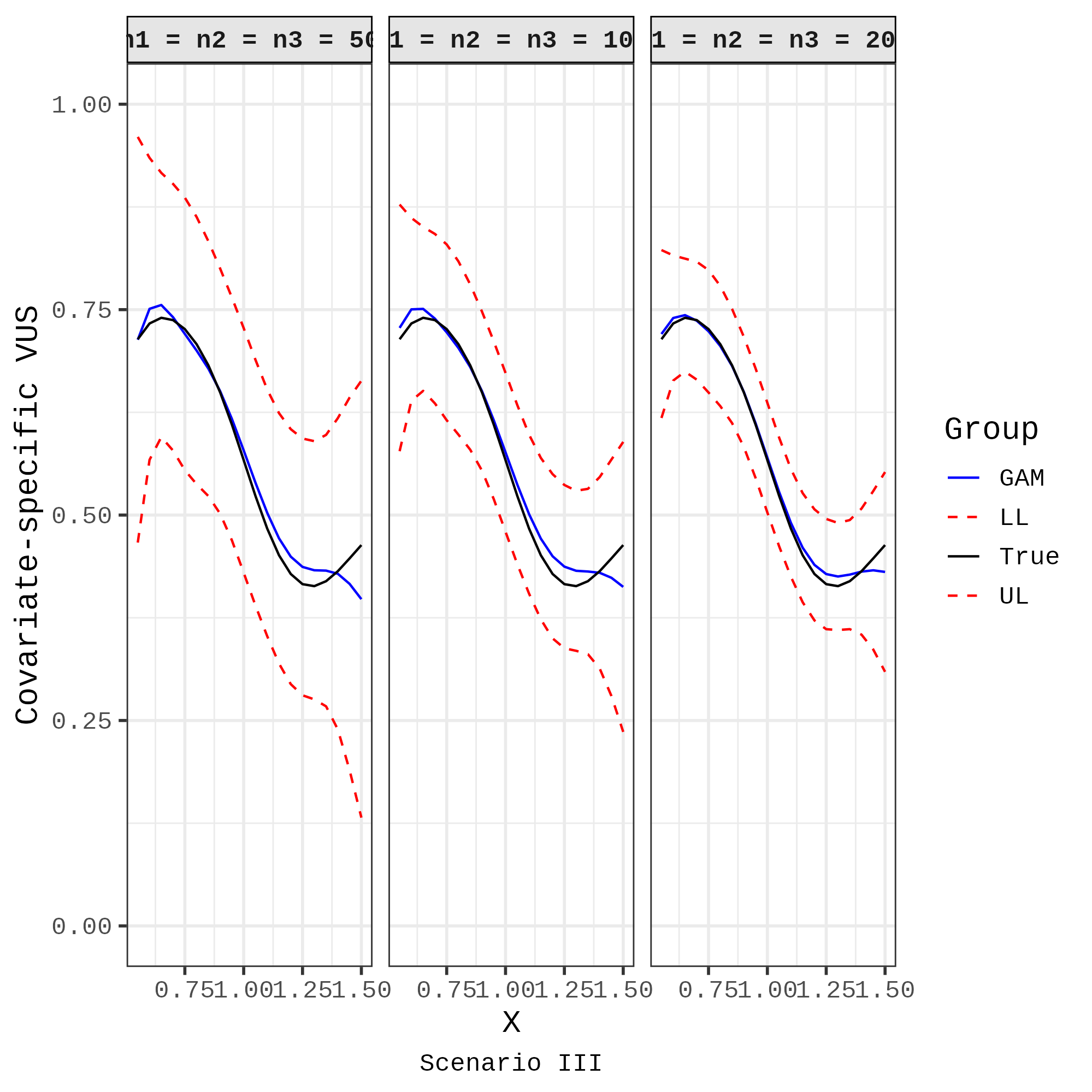}}
\caption{True (black solid line), mean curves (blue solid line) and average
of the 95\% pointwise confidence intervals (red dashed lines) of GAM-based
covariate-specific VUS in scenario I, II and III.}\label{manuscript2026/fig:vus-curves}
\end{figure}

\begin{figure}[htbp]
\centering
\subfloat[Scenario I]{%
  \includegraphics[width=0.48\textwidth,keepaspectratio]{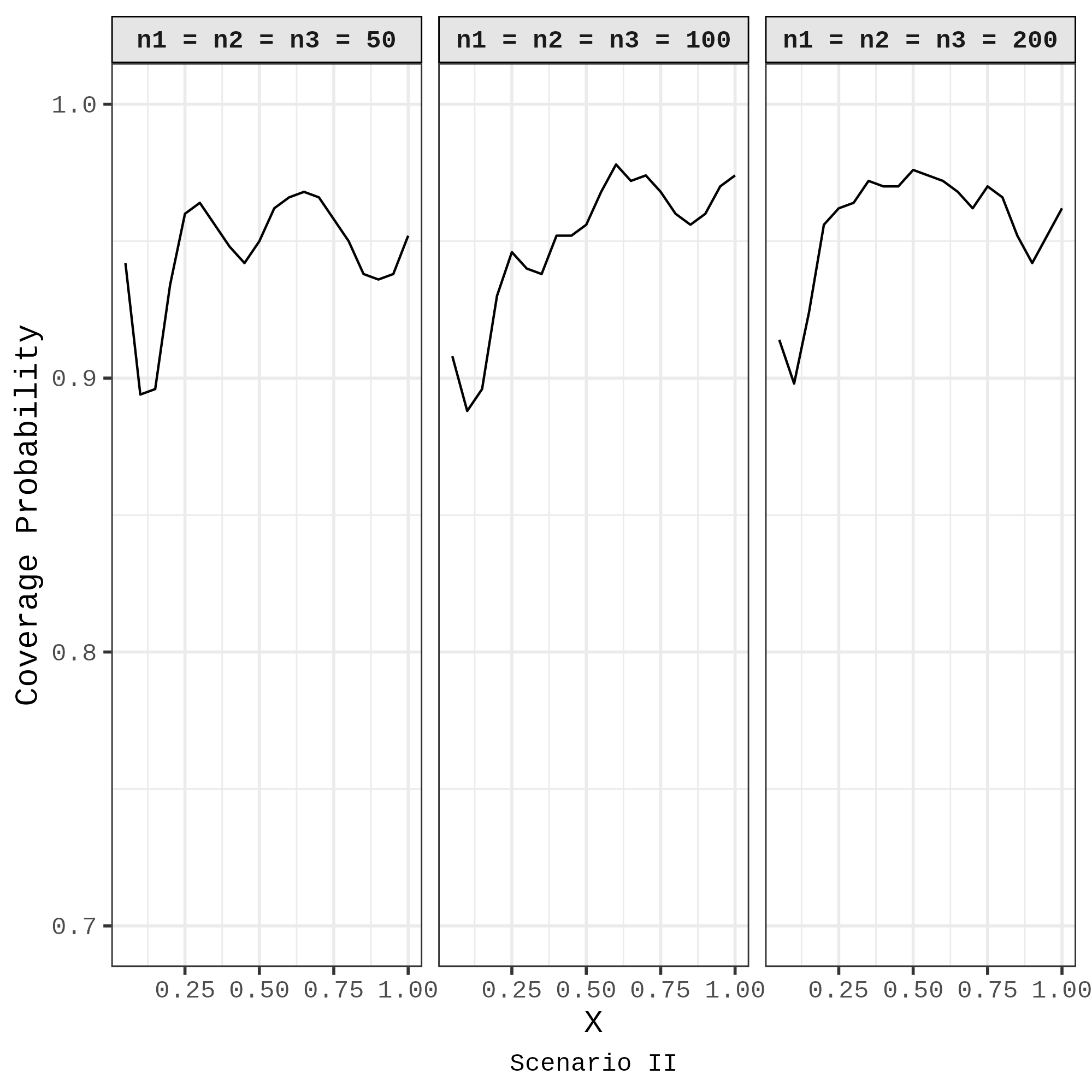}}
\hfill
\subfloat[Scenario II (bsnorm)]{%
  \includegraphics[width=0.48\textwidth,keepaspectratio]{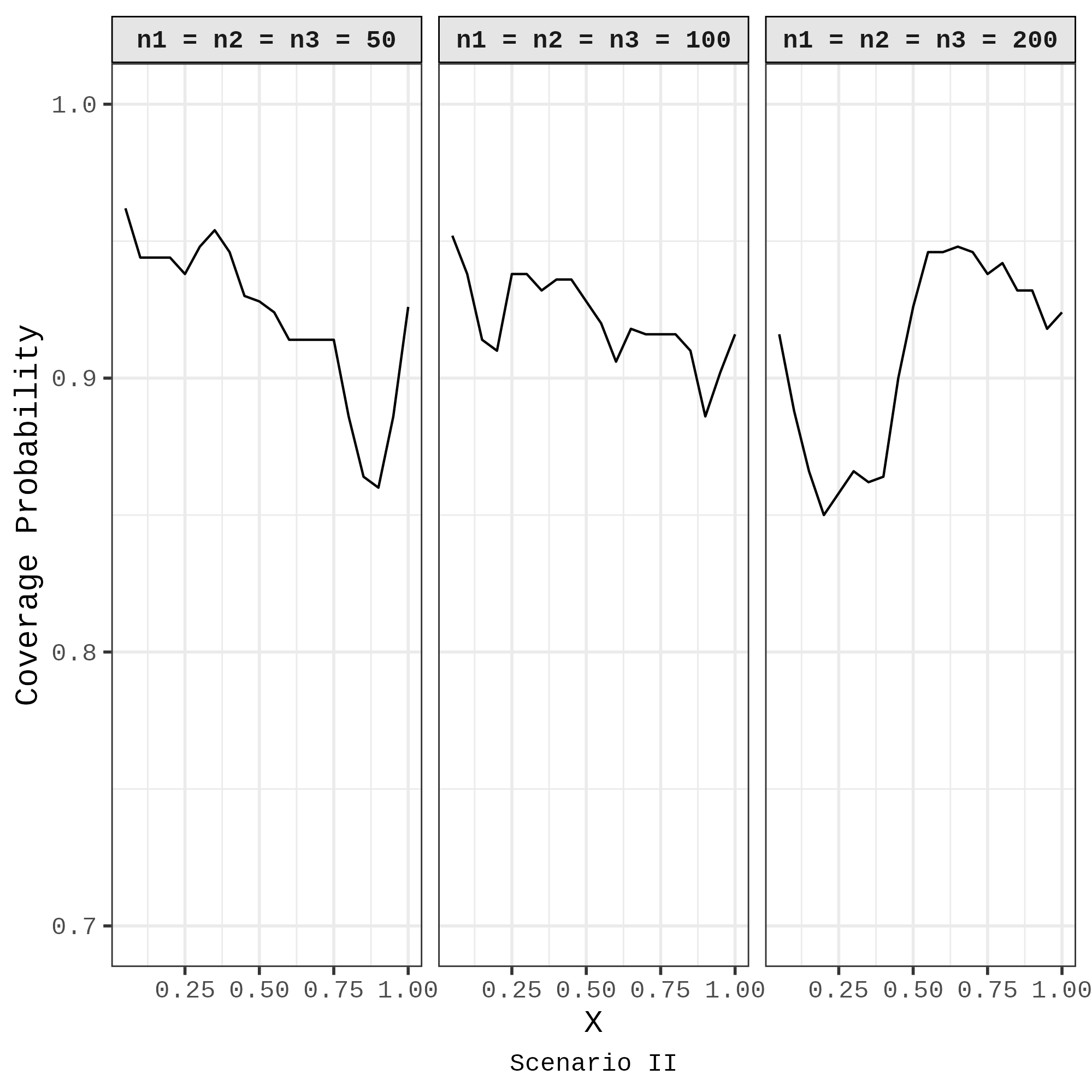}}
\\[0.8em]
\subfloat[Scenario II (gamlss)]{%
  \includegraphics[width=0.48\textwidth,keepaspectratio]{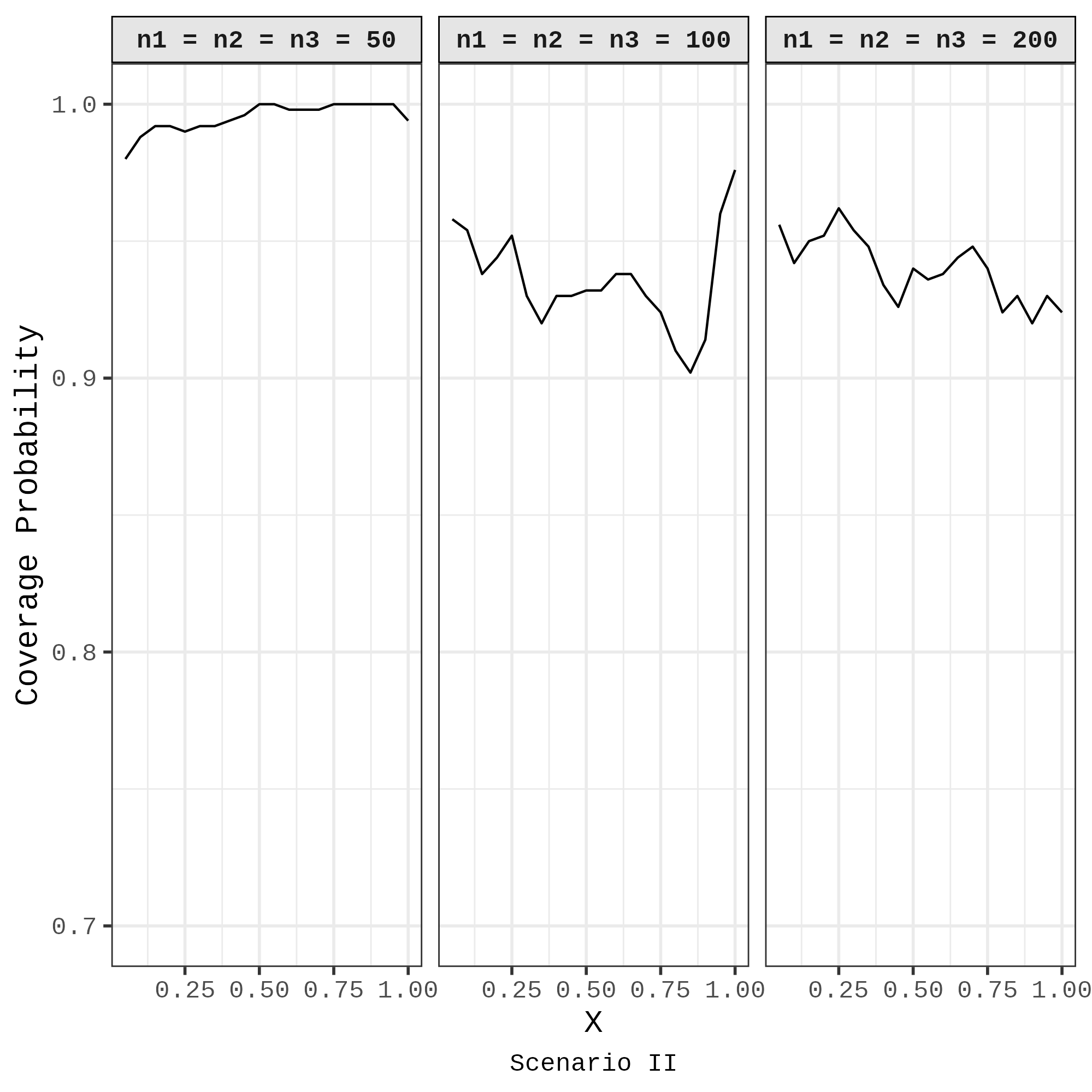}}
\hfill
\subfloat[Scenario III]{%
  \includegraphics[width=0.48\textwidth,keepaspectratio]{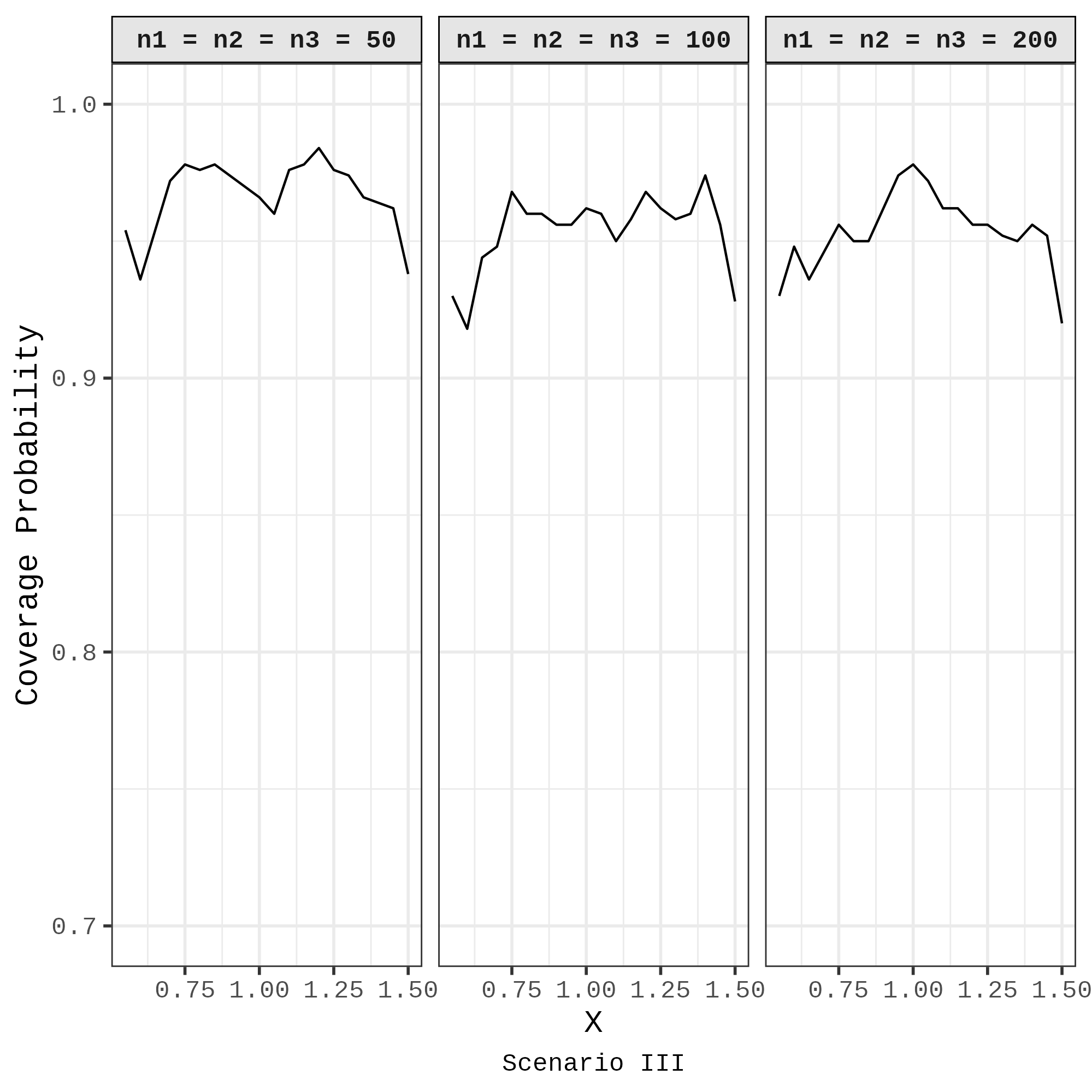}}
\caption{Empirical coverage of GAM-based 95\% pointwise confidence intervals
for covariate-specific VUS in scenario I, II and III.}\label{fig:vus-coverage}
\end{figure}

Figure 1 displays the mean square error (MSE) of GAM-based covariate-specific AUC estimators, at different values $x$ for the covariate. In Scenario I, GAM performs similarly to MW-GEE and MW-LL. When the errors follow heavy-tailed (re-scaled) t-distributions (Scenario II), GAM with t-distribution outperforms MW-GEE and MW-LL in the mid-range of $x$ and have similar MSE around the boundaries. In Scenario III, where the (parametric) working models are misspecified in GEE, GAM-based estimators clearly have lower MSE than both GEE-based estimators across the full range of $x$. In all scenarios, when sample sizes increase, the MSE of GAM-based estimators tends to zero, unlike MW-GEE in the misspecification case. 

The curves representing the averages of the Monte Carlo estimates (MCM) and of the Monte Carlo endpoints of the 95\% pointwise confidence intervals for $x$ are showed in Figure 2 and can be compared to Figures 2-4 in To et al.\cite{to2022estimation} We can see that curves for GAM-based estimators are close to the true ones in all scenarios, and the accuracy improves as the sample sizes increase. The empirical coverage probabilities (CP) for the 95\% confidence intervals are shown in Figures 3 and can be compared to the corresponding results reported in To et al.\cite{to2022estimation} The GAM-based estimators have coverage at or above the nomial level in most cases and are robust in Scenario III where the GEE-based estimators experience undercoverage due to misspecification. Table 1 presents the summary statistics for covariate-adjusted VUS for all scenarios I-III, including Monte Carlo mean (MCM), mean square error
(MSE) multiplied by 100, Monte Carlo standard deviations (MCSD), averages of estimated standard deviations (ASD), and empirical coverages (CP) of the 95\% confidence intervals. The GAM-based estimator performs well in all considered scenarios, comparable to MW-GEE and MW-LL. We observed variance overestimation specific to the t-family specification at $n=50$ which likely reflects small-sample instability in estimating the degrees-of-freedom parameter $\nu$. When $n=100,200$, $\nu$ is estimated more stably, so bootstrap SE and Monte Carlo SD align better.

\begin{table}[h!]
\caption{Simulation results for the covariate-adjusted VUS estimators}
\centering
\small
\begin{tabular}{llccccc}
\toprule
 & & MCM & $\text{MSE} \times 100$ & MCSD & ASD & CP \\
\midrule
\multicolumn{7}{l}{\textit{Scenario I, True AVUS $= 0.691$}} \\
$n = 50$  & GAM   & 0.700 & 0.235 & 0.042 & 0.042 & 0.960\\
          & MW-GEE & 0.682 & 0.219 & 0.046 & 0.046 & 0.954 \\
          & MW-LL  & 0.694 & 0.222 & 0.047 & 0.048 & 0.949 \\ \addlinespace
$n = 100$ & GAM   & 0.696 & 0.124 & 0.030 & 0.030 & 0.954\\
          & MW-GEE & 0.683 & 0.109 & 0.032 & 0.033 & 0.956 \\
          & MW-LL  & 0.690 & 0.105 & 0.032 & 0.033 & 0.955 \\ \addlinespace
$n = 200$ & GAM   & 0.694 & 0.076 & 0.025 & 0.025 & 0.947\\
          & MW-GEE & 0.684 & 0.058 & 0.023 & 0.023 & 0.943 \\
          & MW-LL  & 0.689 & 0.054 & 0.023 & 0.023 & 0.950 \\
\midrule
\multicolumn{7}{l}{\textit{Scenario II, True AVUS $= 0.765$}} \\
$n = 50$  & GAM-norm   & 0.758 & 0.228 & 0.047 & 0.047 & 0.925\\
          & GAM-gamlss &0.737 &0.283 &0.044 &0.086 &1.00\\
          & MW-GEE & 0.742 & 0.338 & 0.053 & 0.051 & 0.919 \\
          & MW-LL  & 0.760 & 0.269 & 0.052 & 0.050 & 0.944 \\ \addlinespace
$n = 100$ & GAM-norm   & 0.762 & 0.133 & 0.034 & 0.034 & 0.935\\
          & GAM-gamlss & 0.778 & 0.094 &0.029 &0.031 &0.930\\
          & MW-GEE & 0.751 & 0.137 & 0.034 & 0.035 & 0.935 \\
          & MW-LL  & 0.762 & 0.115 & 0.034 & 0.034 & 0.956 \\ \addlinespace
$n = 200$ & GAM-norm   & 0.764 & 0.097 & 0.028 & 0.026 & 0.952\\
          & GAM-gamlss & 0.774 & 0.045 & 0.020 & 0.020 &0.900\\
          & MW-GEE & 0.755 & 0.074 & 0.025 & 0.025 & 0.930 \\
          & MW-LL  & 0.761 & 0.063 & 0.025 & 0.024 & 0.948 \\
\midrule
\multicolumn{7}{l}{\textit{Scenario III, True AVUS $= 0.531$}} \\
$n = 50$  & GAM   & 0.538 & 0.238 & 0.038 & 0.035 & 0.956\\
          & MW-GEE & 0.496 & 0.374 & 0.051 & 0.052 & 0.899 \\
          & MW-LL  & 0.534 & 0.270 & 0.052 & 0.053 & 0.954 \\ \addlinespace
$n = 100$ & GAM   & 0.530 & 0.175 & 0.025 & 0.026 & 0.948\\
          & MW-GEE & 0.499 & 0.226 & 0.035 & 0.036 & 0.850 \\
          & MW-LL  & 0.532 & 0.126 & 0.035 & 0.036 & 0.951 \\ \addlinespace
$n = 200$ & GAM   & 0.534 & 0.102 & 0.020 & 0.019 & 0.950\\
          & MW-GEE & 0.500 & 0.161 & 0.026 & 0.026 & 0.752 \\
          & MW-LL  & 0.531 & 0.064 & 0.025 & 0.025 & 0.946 \\
\bottomrule
\end{tabular}
\end{table}

\section{Application to ADNI Data}
In this section, we apply the proposed methods to distinguish the stages of Alzheimer's disease (AD). The data were obtained from the Alzheimer Disease Neuroimaging Initiative (ADNI, adni.loni.usc.edu). The Alzheimer's Disease Neuroimaging Initiative (ADNI) is a multi-center collaborative AD open-source database. ADNI integrates researchers with research data to assess the progression of Mild Cognitive Impairment (MCI) and early AD. The database includes various longitudinal data types such as MRI, PET scans, biomarkers, and clinical neuropsychological assessments. Additionally, the project aims to identify biomarkers for diagnosing and predicting AD, and to develop potential biomarkers for clinical applications. For up-to-date information, see www.adni-info.org.

In our application, we consider 1249 subjects who are registered in three ADNI projects, which are ADNI1, ADNI2, and ADNI GO. Among them, we have 362 Cognitive Normal (CN), 528 Mild Cognitive Impairment (MCI), and 318 AD subjects. Among the biomarkers contained in the project, we consider three of them, which are \(A\beta 1-42\)(amyloid-\(\beta 1\)-42), Tau (total tau) protein, and pTau (phosphorylated tau) protein.

In our analysis, we consider three covariates: age (log-transformed), education and gender. Figure 4 displays the covariate-specific VUS and Table 2 presents the covariate-adjusted VUS. The confidence intervals are based on 1000 bootstrap samples.

\begin{figure}[htbp]
\centering
\includegraphics[height=0.4\textheight, keepaspectratio]{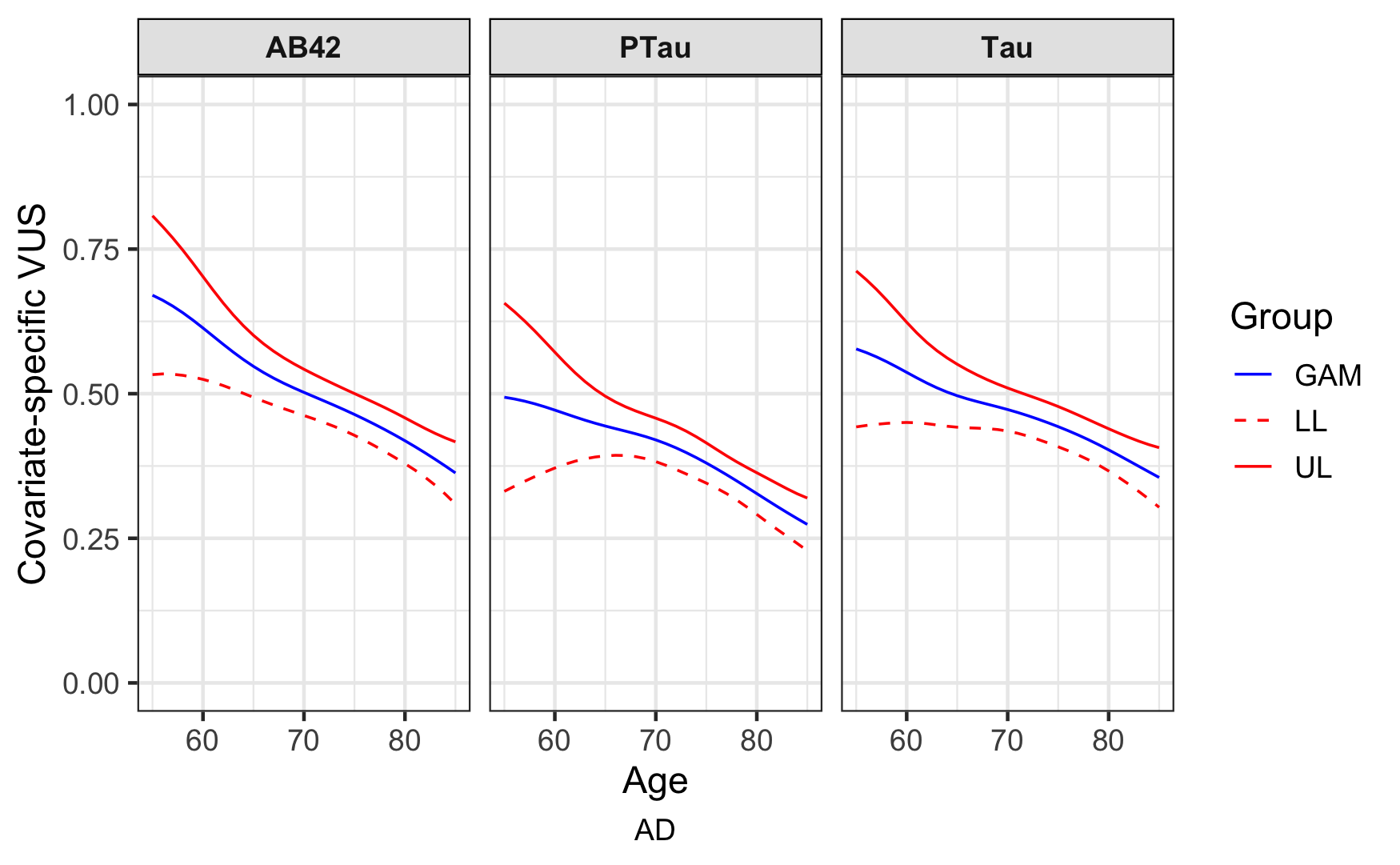}
\caption{Age-specific VUS estimates with 95\% pointwise confidence intervals.}
\end{figure}

\begin{table}
\caption{ Estimated Covariate Adjusted VUS for \(A\beta1-42\), \(Tau\) and \(pTau\).}
\begin{center}
\begin{tabular}{lcccc}
\toprule
Biomarker & Estimate & S.E. & 95\% CI \\
\hline
$A\beta1-42$ & 0.478 & 0.015 & [0.449,0.507]\\
$Tau$ & 0.385 & 0.014 & [0.358,0.413]\\
$pTau$ & 0.449 & 0.015 & [0.420,0.478]\\
\hline
\end{tabular}
\end{center}
\end{table}

The results are quite different than the unadjusted VUS estimates for $A\beta1-42$ (0.378, 95\% confidence interval (0.346, 0.409)), for Tau (0.360, 95\% confidence interval (0.329, 0.391)) and for pTau (0.284, 95\% confidence interval (0.255, 0.312)). Overall, the results indicate that the diagnostic accuracy of $A\beta1-42$ is the highest.

\textcolor{black}{To compare the discriminatory performance of the three biomarkers, we conduct pairwise comparisons of the covariate--adjusted AUC estimates. We use the bootstrap variance--covariance matrix to account for the correlation among the estimates and constructed normal--approximation confidence intervals for the differences. To adjust for multiplicity, we also construct simultaneous confidence intervals based on multivariate normal distribution with correlation estimated from the bootstrap. The covariate--adjusted AUC for \(A\beta 1-42\) is higher than that for pTau (difference = 0.093, simultaneous 95\% CI: 0.054 to 0.131, unadjusted 95\% CI: 0.060 to 0.125) and not statistically significantly different than Tau (difference = 0.029, simultaneous 95\% CI: -0.013 to 0.071, unadjusted 95\% CI: -0.006 to 0.064), while the difference between pTau and Tau is -0.064 (simultaneous 95\% CI: -0.093 to -0.035, unadjusted 95\% CI: -0.088 to -0.039).}

\section{Discussion}
We have proposed GAM-based estimators for covariate-specific and covariate-adjusted AUC/VUS. These estimators allow non-linear functional forms and interactions among covariates while remain efficient and stable. Simulation studies demonstrate the GAM-based estimators outperform GEE-based estimators in multiple scenarios, especially when sample size is small.  

In the real application, we can potentially choose the best data distribution via cross validation using criterion such as Generalized Akaike Information Criterion (GAIC). However, this approach can only compare the fit of different distributions and does not completely rule out misspecification. An alternative approach is to choose a working distribution to fit GAM, then recover the true distribution using residuals from the working samples non-parametrically. This approach can be numerically unstable when sample size is small and may get time-consuming as sample size increases. 



Recent work has explored flexible machine-learning approaches for covariate-adjusted ROC and VUS estimation, including neural-network–based models. Hammouri et al. \cite{hammouri2025rocanalysiscovariateadjustment} proposed to use neural networks to estimate covariate-dependent conditional means and variances of the biomarker within each population. The ROC and VUS are then derived under a normality assumption. While flexible with respect to nonlinear covariate effects, this approach constrains the conditional distributions to be Gaussian, thereby capturing only the first two moments. In contrast, the proposed GAM-based approach allows greater distributional flexibility through smooth modeling and alternative distributional specifications, which can accommodate departures from normality such as skewness, heavy tails, or multimodality. Such flexibility is important because discrimination measures, including the AUC and VUS, depend on the entire conditional distributions of the biomarker rather than solely on their mean and variance, as shown in Eq. (2). Consequently, more flexible distributional modeling may yield more accurate estimates of diagnostic performance when the Gaussian assumption is violated.

Software such as \texttt{ROCnReg} \citep{rodriguez2020rocnreg} is available for covariate adjustment in ROC analysis but incorporation of GAM is still needed. We provide R code implementing the proposed methods in the Supplement and will publish our R package in the future.

\section*{Author Contributions}

Shanshan Liu and Jianlei Huang contributed equally to this work.

\section*{Acknowledgments}
The authors thank Dr. To for sharing the computer program that implemented the methods of To et al. \citep{to2022estimation}.

\section*{Financial Disclosure}

None reported.

\section*{Conflicts of Interest}

The authors declare no conflicts of interest.
\bibliographystyle{plainnat}
\bibliography{GAM_AUC}

\section*{Supporting Information}

Additional Supporting Information may be found in the online version of the article at the publisher's website.  The Supporting Informations include an R code that implements the proposed methods.

\appendix
\renewcommand{\theequation}{\Alph{section}.\arabic{equation}}
\renewcommand{\thetable}{\Alph{section}.\arabic{table}}
\renewcommand{\thefigure}{\Alph{section}.\arabic{figure}}
\setcounter{equation}{0}
\setcounter{table}{0}
\setcounter{figure}{0}

\section*{Appendix}
\section{Asymptotic Distribution Theory}

\subsection{General framework}

We view the proposed estimators as sample averages of smooth
functionals involving group-specific nuisance quantities. This mirrors the logic of
Janes and Pepe (2009), who derive asymptotic theory for covariate-adjusted ROC functionals
by combining (i) asymptotic linearity of nuisance estimators with (ii) an empirical averaging step.

Throughout the Appendix we write $Z=(Y,D,\bm X)$ for the observed data on a single subject and $Z_i=(Y_i,D_i,\bm X_i)$ for the $i$th replicate, and we retain the notation of Section~2: $F_k(\cdot\mid\bm x)$ denotes the conditional c.d.f. of the test result in disease group $k$, $\bm\theta_k(\bm x)$ the corresponding vector of distributional parameters, and $H$ the c.d.f. of $\bm X$.

For binary $D\in\{0,1\}$, define the map
\[
\mathcal T_{\bm x}(F_0,F_1)=\int F_0(y\mid \bm x)\,dF_1(y\mid \bm x),
\]
so that $\mathrm{AUC}(\bm x)=\mathcal T_{\bm x}(F_0,F_1)$ and
$\widehat{\mathrm{AUC}}(\bm x)=\mathcal T_{\bm x}(\widehat F_0,\widehat F_1)$.
For three-class ordered $D\in\{1,2,3\}$, define
\[
\mathcal V_{\bm x}(F_1,F_2,F_3)=P(Y_1<Y_2<Y_3\mid \bm X=\bm x),
\]
so that $\mathrm{VUS}(\bm x)=\mathcal V_{\bm x}(F_1,F_2,F_3)$ and
$\widehat{\mathrm{VUS}}(\bm x)=\mathcal V_{\bm x}(\widehat F_1,\widehat F_2,\widehat F_3)$.

\section{Assumptions}
Besides the assumption on the covariance matrix of X, we impose the following assumptions.

\textbf{(A1)} Observations are independently sampled conditional on disease
status $D$. As $n\to\infty$, the group sample sizes satisfy $n_k\to\infty$
and $n_k/n \to \pi_k \in (0,1)$ for all disease groups $k$.

\textbf{(A2)} For each group $k$ and each fixed $\bm x \in \mathcal X$, the
distributional GAM estimator $\widehat{\bm\theta}_k(\bm x)$ is consistent and
asymptotically normal:
\[
\sqrt{n_k}\{\widehat{\bm\theta}_k(\bm x)-\bm\theta_k(\bm x)\}
\Rightarrow N\{\bm 0,\bm\Sigma_k(\bm x)\}.
\]

\textbf{(A3)} For each $y$ and $\bm x$, the CDF $F_{\mathcal D}(y;\bm \theta)$ is continuously
differentiable with respect to $\bm \theta$ in a neighborhood of $\bm \theta_k(\bm x)$,
and the derivative $\dot F_D(y;\bm\theta)=\partial F_D(y;\bm\theta)/\partial\bm\theta$ is bounded in $y$.

\textbf{(A4)} The conditional densities corresponding to
$F_k(\cdot\mid \bm x)$ exist, are bounded, and are uniformly positive on the
support of $Y$.

Assumption (A1) reflects a common sampling framework in diagnostic studies and ensures that each disease group contributes non-negligible information.
Assumption (A2) provides the asymptotic linearity required to derive the large-sample distribution of the proposed estimators; penalized likelihood estimators in generalized additive models satisfy this property under standard regularity conditions when smoothing parameters are estimated by REML or related criteria \citep{wood2016smoothing,wood2017generalized}. Assumption (A3) guarantees that small perturbations in the estimated parameters induce smooth perturbations of the conditional distributions, allowing linearization of the ROC functionals. Assumption (A4) ensures that the Mann–Whitney representations of AUC and VUS are well defined. 

\section{Proofs}

\subsection{Proof of Theorem 1}

Fix $\bm x\in\mathcal X$. The proposed estimator of the covariate-specific AUC is obtained by substituting the estimated conditional distributions
$\widehat F_k(\cdot\mid \bm x)$ into the functional
$\mathcal T_{\bm x}(F_0,F_1)$.

Under Assumption (A2), the parameter estimators admit asymptotic linear representations. By smoothness of the distributional model,
\[
\widehat F_k(y\mid \bm x)-F_k(y\mid \bm x)
=
\dot F_D\!\left(y;\bm\theta_k(\bm x)\right)^\top
\{\widehat{\bm\theta}_k(\bm x)-\bm\theta_k(\bm x)\}
+o_p(n_k^{-1/2}),
\]
which yields
\[
\sqrt{n_k}\{\widehat F_k(y\mid \bm x)-F_k(y\mid \bm x)\}
=
\frac{1}{\sqrt{n_k}}\sum_{i:D_i=k}\phi_{k,x}(y;Z_i)+o_p(1),
\]
where
\[
\phi_{k,\bm x}(y;Z)
=
\dot F_D\!\left(y;\bm\theta_k(\bm x)\right)^\top \bm\psi_{k,x}(Z),
\]
and $\bm\psi_{k,x}$ is the influence function of
$\widehat{\bm\theta}_k(\bm x)$.

Substituting into the AUC functional and retaining first-order terms gives
\[
\widehat{\mathrm{AUC}}(\bm x)-\mathrm{AUC}(\bm x)
=
\int\{\widehat F_0-F_0\}\,dF_1
+
\int F_0\,d(\widehat F_1-F_1)
+o_p(n^{-1/2}).
\]

Using the linear representations above,
\[
\int\{\widehat F_0-F_0\}\,dF_1
=
\frac{1}{n_0}\sum_{i:D_i=0}\xi_{0,x}(Z_i)+o_p(n_0^{-1/2}),
\]
\[
\int F_0\,d(\widehat F_1-F_1)
=
\frac{1}{n_1}\sum_{i:D_i=1}\xi_{1,x}(Z_i)+o_p(n_1^{-1/2}),
\]
where the functions $\xi_{k,\bm x}$ are defined by
\[
\xi_{0,\bm x}(Z)
=
\int \phi_{0,\bm x}(y;Z)\, dF_1(y\mid \bm x),
\qquad
\xi_{1,\bm x}(Z)
=
\int F_0(y\mid \bm x)\, d\phi_{1,\bm x}(y;Z).
\]

Therefore,
\[
\sqrt{n}\{\widehat{\mathrm{AUC}}(\bm x)-\mathrm{AUC}(\bm x)\}
=
\frac{1}{\pi_0}\cdot \frac{1}{\sqrt{n_0}}\sum_{i:D_i=0}\xi_{0,\bm x}(Z_i)
+
\frac{1}{\pi_1}\cdot \frac{1}{\sqrt{n_1}}\sum_{i:D_i=1}\xi_{1,\bm x}(Z_i)
+o_p(1).
\]

Since the sums are taken over disjoint disease groups,
they are independent under the sampling scheme in Assumption (A1). The central limit theorem implies that the right-hand side converges to a mean-zero normal distribution with variance
\[
\sigma^2_{\mathrm{AUC}}(\bm x)
=
\pi_0^{-1}Var\{\xi_{0,\bm x}(Z)\mid D=0\}
+
\pi_1^{-1}Var\{\xi_{1,\bm x}(Z)\mid D=1\}.
\]
This establishes asymptotic normality of $\widehat{\mathrm{AUC}}(\bm x)$.
\hfill$\square$

\subsection{Proof of Theorem 2}

By definition of the empirical distribution $\widehat H$,
\[
\widehat{\mathrm{AAUC}}
=
\int \widehat{\mathrm{AUC}}(\bm x)\,d\widehat H(\bm x)
=
\frac{1}{n}\sum_{i=1}^n \widehat{\mathrm{AUC}}(\bm X_i),
\qquad
\mathrm{AAUC}=E\{\mathrm{AUC}(\bm X)\}.
\]
Add and subtract $\mathrm{AUC}(\bm X_i)$ to obtain
\begin{equation}
\widehat{\mathrm{AAUC}}-\mathrm{AAUC}
=
\underbrace{\frac{1}{n}\sum_{i=1}^n\{\mathrm{AUC}(\bm X_i)-\mathrm{AAUC}\}}_{(I)}
+
\underbrace{\frac{1}{n}\sum_{i=1}^n\{\widehat{\mathrm{AUC}}(\bm X_i)-\mathrm{AUC}(\bm X_i)\}}_{(II)}.
\label{eq:AAUC_decomp_JP}
\end{equation}

Term (I) is an average of i.i.d.\ mean-zero variables with finite variance, hence
\[
\sqrt{n}\,(I)
=
\frac{1}{\sqrt{n}}\sum_{i=1}^n\{\mathrm{AUC}(\bm X_i)-\mathrm{AAUC}\}
\Rightarrow
N\!\left(0,\mathrm{Var}\{\mathrm{AUC}(\bm X)\}\right).
\]

For term (II), apply the first-order expansion from Theorem~1:
for each fixed $\bm x\in\mathcal X$,
\begin{equation}
\widehat{\mathrm{AUC}}(\bm x)-\mathrm{AUC}(\bm x)
=
\frac{1}{n_0}\sum_{j:D_j=0}\xi_{0,\bm x}(Z_j)
+
\frac{1}{n_1}\sum_{j:D_j=1}\xi_{1,\bm x}(Z_j)
+ o_p(n^{-1/2}),
\label{eq:AUC_lin_JP}
\end{equation}
with $E\{\xi_{k,\bm x}(Z)\mid D=k\}=0$.
Evaluating \eqref{eq:AUC_lin_JP} at $\bm x=\bm X_i$ and averaging over $i$ yields
\[
(II)
=
\frac{1}{n_0}\sum_{j:D_j=0}\left\{\frac{1}{n}\sum_{i=1}^n \xi_{0,\bm X_i}(Z_j)\right\}
+
\frac{1}{n_1}\sum_{j:D_j=1}\left\{\frac{1}{n}\sum_{i=1}^n \xi_{1,\bm X_i}(Z_j)\right\}
+ o_p(n^{-1/2}).
\]
Let $\eta_k(Z)=E\{\xi_{k,\bm x}(Z)\}$, where $\bm X\sim H$ is independent of $Z$.
By the law of large numbers, for each $j$,
$n^{-1}\sum_{i=1}^n \xi_{k,\bm X_i}(Z_j)\xrightarrow{p}\eta_k(Z_j)$, and therefore
\[
\sqrt{n}\,(II)
=
\frac{\sqrt{n}}{n_0}\sum_{j:D_j=0}\eta_0(Z_j)
+
\frac{\sqrt{n}}{n_1}\sum_{j:D_j=1}\eta_1(Z_j)
+o_p(1).
\]
Using $n_k/n\to\pi_k$ (Assumption (A1)), we may write
\[
\sqrt{n}\,(II)
=
\frac{1}{\pi_0}\cdot \frac{1}{\sqrt{n_0}}\sum_{j:D_j=0}\eta_0(Z_j)
+
\frac{1}{\pi_1}\cdot \frac{1}{\sqrt{n_1}}\sum_{j:D_j=1}\eta_1(Z_j)
+o_p(1).
\]
A central limit theorem applied within each disease group implies that the two
terms converge jointly to mean-zero normal limits. Combining the limits for (I)
and (II) in \eqref{eq:AAUC_decomp_JP} and applying Slutsky's theorem gives
\[
\sqrt{n}\{\widehat{\mathrm{AAUC}}-\mathrm{AAUC}\}\Rightarrow N(0,\sigma^2_{\mathrm{AAUC}}),
\]
where $\sigma^2_{\mathrm{AAUC}}$ is the variance of the limiting sum of the
component terms. This establishes the result for AAUC. \hfill$\square$

\subsection{VUS and adjusted VUS}

The arguments for $\widehat{\mathrm{VUS}}(\bm x)$ and
$\widehat{\mathrm{AVUS}}$ follow analogously, replacing the functional $\mathcal T_{\bm x}$ by $\mathcal V_{\bm x}$ and applying the same linearization to the multivariate mapping $(F_1,F_2,F_3)\mapsto \mathcal V_{\bm x}(F_1,F_2,F_3)$.
Under Assumptions A1--A4,
\[
\sqrt{n}\{\widehat{\mathrm{VUS}}(\bm x)-\mathrm{VUS}(\bm x)\}\Rightarrow N\{0,\sigma^2_{\mathrm{VUS}}(\bm x)\},
\qquad
\sqrt{n}\{\widehat{\mathrm{AVUS}}-\mathrm{AVUS}\}\Rightarrow N(0,\sigma^2_{\mathrm{AVUS}}),
\]
with variance decompositions analogous to those for AUC and AAUC.

\nocite{*}

\end{document}